\documentclass[11pt]{article}

\usepackage[margin=0.95in]{geometry}
\usepackage{amsmath,amssymb,amsthm}
\usepackage{graphicx}
\usepackage{booktabs}
\usepackage{array}
\usepackage{multirow}
\usepackage{caption}
\usepackage{authblk}
\usepackage[colorlinks=true,citecolor=black,linkcolor=black,urlcolor=black]{hyperref}
\usepackage[round]{natbib}
\usepackage{enumitem}
\setlist[itemize]{topsep=2pt,itemsep=2pt,parsep=0pt}

\title{\textbf{Evaluation of Circular Logistic Regression Models with Asymmetric Link Functions}}
\author{Feridun Tasdan}
\affil{Department of Mathematics, Western Illinois University, Macomb, IL, USA \\ \texttt{f-tasdan@wiu.edu}}
\date{}

\begin{document}
\maketitle

\begin{abstract}
\noindent Circular (directional) data arise whenever observations are measured as angles on the unit circle, such as wind direction, time of day, or calendar phase, and require statistical methods that respect the periodicity of the domain $[0,2\pi)$. While circular-linear and linear-circular regression models are well established, regression models for a binary or binomial response observed jointly with a circular predictor remain largely undeveloped, with the sole closely related study restricted to the symmetric logit link. This paper develops and evaluates a circular logistic regression framework in which the linear predictor is expressed through the cosine and sine of the circular covariate, and compares the performance of symmetric link functions (logit, probit) against asymmetric alternatives (complementary log-log, Cauchit, and a skew-logit power link) under a generalized linear model formulation. A Monte Carlo simulation generates circular predictors from the von Mises distribution under two concentration regimes and evaluates model fit using the Akaike Information Criterion (AIC) and deviance. The methodology is illustrated with two real data sets: daily rainfall occurrence and wind direction recorded in Macomb, Illinois, and monthly earthquake counts in Western Anatolia, Turkiye, the latter used to connect the binary circular model to the related circular Poisson regression framework for count outcomes. Results indicate that the choice of link function matters most when the circular predictor is broadly dispersed and the response is markedly unbalanced; under high concentration of the predictor, symmetric links are preferred and asymmetric links are prone to instability. Practical guidelines and directions for future software development are discussed.

\end{abstract}

\section{Introduction}

Most statistical regression methods assume that both the response and the predictor variables are measured on the real line. A substantial number of applications, however, involve variables recorded as directions or angles such as wind direction in meteorology, the time of day or day of year an event occurs, migratory headings of animals, or the orientation of geological features. Such circular (or directional) data live on the unit circle with domain $[0,2\pi)$ in radians, and the point $0$ is identified with $2\pi$. Because the origin of measurement is arbitrary and the scale wraps around, ordinary linear model technique applied directly to angles produces misleading summaries and inferences; specialized circular statistical methods are required \citep{fisher1993,mardia2009,jammalamadaka2001,pewsey2013}.

Regression methodology for circular data has developed along two main lines: circular-linear regression, in which a circular response is modeled as a function of linear predictors, and linear-circular regression, in which a linear response is modeled as a function of a circular predictor \citep{fisherlee1992,jammalamadaka2001}. The linear-circular case is typically handled by re-expressing the circular covariate $\theta$ through its cosine and sine components, which converts the problem into an ordinary linear regression in $\cos(\theta)$ and $\sin(\theta)$ \citep{jammalamadaka2001}. This cosine--sine device is the foundation of the present paper.

Far less attention has been paid to the case where the response variable is binary or binomial (success/failure, presence/absence, occurrence/non-occurrence) and the predictor is a circular measurement. To the best of our knowledge, the only closely related study is \citet{kadhem2017}, who outlined a circular logistic regression model restricted to the canonical logit link. \citet{tasdan2026} recently extended the circular generalized linear model idea to count responses by developing a circular Poisson regression model for data such as the number of earthquakes occurring within each month of the year, and noted explicitly that no parallel treatment of circular logistic regression with a comparison of link functions existed in the literature. Independently, in the non-circular setting, \citet{dawotola2025} compared symmetric and asymmetric link functions (logit, probit, complementary log-log, Laplace, and Cauchy) for binomial regression applied to bioassay data, building on a substantial methodological literature establishing that the choice of link function is not a minor implementation detail: \citet{czado1992} showed that link misspecification can produce substantial bias and inflated mean squared error in both the regression coefficients and the fitted success probabilities; \citet{chen1999} proposed a skewed link family and argued that asymmetric links are preferable when the binary response is markedly unbalanced; \citet{collett2003} reached a similar conclusion in applied binary-data modeling; and \citet{gomezdeniz2022} introduced a new asymmetric logit-type link motivated by unbalanced binary outcomes common in insurance and economic decision data.

This paper brings these two strands of literature together. Related circular regression methodology includes general angular regression models for animal-movement ecology \citep{rivest2015} and recent guidance on comparing independent samples of circular data \citep{landler2021}, underscoring that circular regression methods continue to be actively developed across application domains. We (i) formally develop a circular logistic regression model in which the linear predictor is a cosine-sine function of a circular covariate, paralleling the linear-circular regression model and the circular Poisson regression model of \citet{tasdan2026}; (ii) review and assemble a working catalogue of symmetric and asymmetric link functions suitable for this setting, including the logit, probit, complementary log-log, Cauchit, and a power (skew-logit) link; (iii) conduct a Monte Carlo simulation study, generating the circular predictor from the von Mises distribution under two concentration regimes, to evaluate how link choice interacts with the dispersion of the circular predictor; and (iv) illustrate the methodology with two real data examples of wind direction and rainfall occurrence recorded in Macomb, Illinois, and the Western Anatolia earthquake counts of \citet{tasdan2026}, used here to connect the binary circular model to the circular count-data model. We close with practical recommendations and directions for future software development, since, as of this writing, none of the major circular statistics packages in R \citep{agostinelli2017,hornik2014,pewsey2013} provide a built-in routine for circular logistic regression with a choice of link function.

\section{Essential Concepts of Circular Data}

This section summarizes, briefly, the minimal circular-statistics background needed for the regression models that follow; a full treatment is available in \citet{fisher1993}, \citet{mardia2009}, and \citet{jammalamadaka2001}. Related methodological work on inference procedures for circular data, including multi-sample testing and the handling of tied observations, is discussed in \citet{tasdan2014}, \citet{tasdan2013}, and \citet{tasdan2018}.

Let $\theta_1,\ldots,\theta_n$ denote $n$ angular observations on $[0,2\pi)$. Degrees are converted to radians by multiplying by $\pi/180$, and a time of day or day of year (period $k$) is converted to an angle by $a = 360\cdot X/k$, where $X$ is the elapsed time and $k$ is the full period (e.g., $k=24$ hours or $k=12$ months).

Writing $C = \sum_i\cos(\theta_i)$ and $S = \sum_i\sin(\theta_i)$, the mean resultant length is $R = \sqrt{C^2+S^2}$, with mean resultant length $\bar{R} = R/n \in [0,1]$, and the mean direction is
\begin{equation}
\bar\theta = \arctan(S/C) + I(C<0)\,\pi + I(S<0,\,C>0)\,2\pi. \label{eq:meandir}
\end{equation}
The quantity $V = 1-\bar R$ serves as a circular analogue of variance, bounded on $[0,1]$. The most commonly used circular probability model is the von Mises distribution, $CN(\mu,\kappa)$, with density
\begin{equation}
f(\theta;\mu,\kappa) = \frac{1}{2\pi I_0(\kappa)}\exp\{\kappa\cos(\theta-\mu)\}, \qquad 0\le\theta<2\pi, \label{eq:vonmises}
\end{equation}
where $\mu$ is the mean direction, $\kappa\ge 0$ is the concentration parameter, and $I_0(\kappa)$ is the modified Bessel function of order zero. Larger $\kappa$ produces angles tightly clustered around $\mu$, while $\kappa=0$ reduces to the circular uniform distribution. The von Mises distribution plays the same role in circular statistics that the normal distribution plays on the real line, and it is used below to generate the circular predictor in the simulation study.

\section{Generalized Linear Models for Binary Responses and the Role of the Link Function}

Let $Z_1,\ldots,Z_n$ be independent Bernoulli trials with $P(Z_i=1)=\pi_i$, and let $Y_i=\sum_j Z_j$ denote the number of successes obtained from $n_i$ repeated trials in stratum $i$, so that $Y_i \sim \text{Binomial}(n_i,\pi_i)$. Within the generalized linear model (GLM) framework, the predictor variables enter the model through a linear combination $\eta_i = x_i^T\beta$, and a link function $g$ connects $\eta_i$ to the mean response:
\begin{equation}
g(\pi_i) = x_i^T\beta, \qquad \text{equivalently} \qquad \pi_i = g^{-1}(x_i^T\beta). \label{eq:link}
\end{equation}
Any continuous, strictly increasing function mapping $(-\infty,\infty)$ onto $(0,1)$ is admissible as $g^{-1}$; in practice $g^{-1}$ is usually taken to be the cumulative distribution function (cdf) of a continuous random variable \citep{mccullagh1989,agresti2002}. Table~\ref{tab:links} lists the link/tolerance pairs used in this paper. The logit link corresponds to the logistic cdf and is the canonical link for the Bernoulli/binomial family; it is popular because the resulting coefficients are directly interpretable as log odds ratios \citep{breen2018}. The probit link uses the standard normal cdf $\Phi$. Both logit and probit are symmetric about $\pi=1/2$, in the sense that $g^{-1}(-x) = 1-g^{-1}(x)$.

Symmetric links implicitly assume that $\pi_i$ approaches 0 at the same rate that it approaches 1 as the linear predictor varies. When the observed response is markedly unbalanced which means it contains far more 0s than 1s, or vice versa. Under the unbalance response variable ratio, the symmetrical link function assumption can be inappropriate, and asymmetric (skewed) links often fit better \citep{chen1999,collett2003,gomezdeniz2022}. Practical guidance on choosing among candidate links more generally is reviewed by \citet{li2014}, and the theoretical justification for the logistic function itself in binary regression is examined by \citet{zaidi2023}. The complementary log-log (cloglog) link, derived from the Gumbel (extreme value) cdf, approaches 1 more quickly than it approaches 0 and is a standard asymmetric alternative, widely used in dose-response and survival applications \citep{dawotola2025,shim2023}. The Cauchit link, based on the heavy-tailed Cauchy cdf, remains symmetric but is far less sensitive to extreme covariate values than logit or probit. Finally, skew (power) families generalize the logit by introducing an additional shape parameter $\sigma$ that controls the rate of approach to the boundaries; the skew-logit link used in this paper raises the logistic cdf to a power $\sigma$ estimated jointly with the regression coefficients, while related skew-Laplace and skew-Cauchy families are obtained analogously from the Laplace and Cauchy distributions \citep{dawotola2025,smithson2017}.

\begin{table}[ht]
\centering
\caption{Link functions used in this study: regression ($g$) and tolerance ($g^{-1}$) forms.}
\label{tab:links}
\small
\begin{tabular}{lll}
\toprule
\textbf{Link} & \textbf{Regression model} $g(\pi_i)$ & \textbf{Tolerance model} $\pi_i = g^{-1}(\eta_i)$ \\
\midrule
Logit & $\log[\pi_i/(1-\pi_i)] = \eta_i$ & $\exp(\eta_i)/[1+\exp(\eta_i)]$ \\
Probit & $\Phi^{-1}(\pi_i) = \eta_i$ & $\Phi(\eta_i)$ \\
Cloglog & $\log[-\log(1-\pi_i)] = \eta_i$ & $1-\exp[-\exp(\eta_i)]$ \\
Cauchit & $\tan[\pi(\pi_i-1/2)] = \eta_i$ & $(1/\pi)\arctan(\eta_i) + 1/2$ \\
Skew-logit ($\sigma$) & $\log[\pi_i^{1/\sigma}/(1-\pi_i^{1/\sigma})] = \eta_i$ & $[\exp(\eta_i)/(1+\exp(\eta_i))]^{\sigma}$ \\
\bottomrule
\end{tabular}
\end{table}

Here $\eta_i = x_i^T\beta$ is the linear predictor and $\Phi$ denotes the standard normal cdf. The skew-logit tolerance model reduces exactly to the ordinary logit when $\sigma=1$, and the additional parameter $\sigma$ is estimated jointly with $\beta$ via maximum likelihood, exactly as in the bioassay-data application of \citet{dawotola2025}.

Given observed data, the regression coefficients are obtained by maximum likelihood. Writing the likelihood in terms of the tolerance function $F=g^{-1}$,
\begin{equation}
L(\beta\mid y,x) = \prod_i [F(\eta_i)]^{y_i}[1-F(\eta_i)]^{1-y_i}, \label{eq:likelihood}
\end{equation}
the log-likelihood is maximized numerically (Newton--Raphson or iteratively reweighted least squares for the standard links; direct numerical optimization for the skew links), since closed-form solutions are generally unavailable \citep{mccullagh1989}.

\section{Linear-Circular Regression as a Building Block}

Before introducing the circular logistic model, it is useful to recall the classical linear-circular regression model, in which a linear response $Y$ is regressed on a circular predictor $\theta$ \citep{jammalamadaka2001}:
\begin{equation}
E[Y\mid\theta_i] = \beta_0 + A\cos(\theta_i-\theta_0). \label{eq:lincirc}
\end{equation}
Here $\beta_0$ is the expected response when $\theta_i-\theta_0 = 90^\circ$, $A$ is the amplitude of the cyclic fluctuation in the response, and $\theta_0$ is the acrophase, which the angle that the response attains its maximum. Expanding the cosine of a difference gives the equivalent, estimable form
\begin{equation}
E[Y\mid\theta_i] = \beta_0 + \beta_1\cos(\theta_i) + \beta_2\sin(\theta_i), \label{eq:lincirc2}
\end{equation}
with $A=\sqrt{\beta_1^2+\beta_2^2}$ and $\theta_0=\arctan(\beta_2/\beta_1)$ (quadrant-adjusted). Equation~\eqref{eq:lincirc2} is linear in $\cos(\theta)$ and $\sin(\theta)$ and can therefore be fit with ordinary least squares once $\theta$ has been transformed; an entirely analogous device underlies the circular Poisson regression model of \citet{tasdan2026}, where the same cosine-sine linear predictor is exponentiated to keep the Poisson mean positive. The circular logistic model developed next applies the identical transformation, but maps the cos--sin linear predictor through a link function appropriate for a Bernoulli/binomial response rather than a Gaussian or Poisson one.

\section{Circular Logistic Regression Model}

Consider $n$ independent Bernoulli observations $Z_i$ with success probability $\pi_i$, where $\pi_i$ depends on a circular predictor $\theta_i$. We define the circular logistic regression model by
\begin{equation}
g(\pi_i) = \beta_0 + A\cos(\theta_i-\theta_0), \label{eq:circlogit}
\end{equation}
where $g$ is one of the link functions of Table~\ref{tab:links}. As in Section~4, $\beta_0$ is the value of the link function (e.g., the log-odds, under the logit link) when $\theta_i-\theta_0=90^\circ$, $A$ is the amplitude of the cyclic fluctuation in the linked response, and $\theta_0$ is the angle at which the link function $g$ attains its maximum (hence $\pi_i$) since $g$ is monotone funcyion. Expanding as before yields the streamlined, directly fittable form
\begin{equation}
g(\pi_i) = \beta_0 + \beta_1\cos(\theta_i) + \beta_2\sin(\theta_i) = \eta_i, \label{eq:circlogit2}
\end{equation}
with $A=\sqrt{\beta_1^2+\beta_2^2}$ and $\theta_0=\text{atan2}(\beta_2,\beta_1)$. Equation~\eqref{eq:circlogit2} is an ordinary GLM with two engineered covariates, $\cos(\theta_i)$ and $\sin(\theta_i)$, so that any link function compatible with the binomial family like logit, probit, cloglog, Cauchit, or a skew-logit power link can in principle be substituted without altering the cosine-sine parameterization. This is the key practical advantage of the cosine-sine model: it converts a circular predictor problem into a standard two-covariate GLM that existing software can fit directly, while the amplitude $A$ and phase angle $\theta_0$ recovered from $\beta_1$ and $\beta_2$ retain a direct circular interpretation. Maximum likelihood estimates of $(\beta_0,\beta_1,\beta_2,[\sigma])$ are obtained by maximizing the likelihood in Equation~\eqref{eq:likelihood} with $\eta_i$ given by Equation~\eqref{eq:circlogit2}, exactly as in standard binomial GLM estimation.

\section{Simulation Study}

A Monte Carlo simulation was conducted to evaluate how the five link functions of Table~\ref{tab:links} (logit, probit, cloglog, Cauchit, and skew-logit) compare under the circular logistic model of Equation~\eqref{eq:circlogit2}, and to see how this comparison is affected by the concentration of the circular predictor around its mean direction.

\subsection*{Design}
\begin{itemize}
\item Sample size $n=100$ per replication.
\item Circular predictor $\theta_i$ generated from the von Mises distribution, $CN(\mu=\pi/2,\kappa)$, under two concentration regimes: $\kappa=3$ (a dispersed predictor, spread broadly around $\pi/2$) and $\kappa=12$ (a concentrated predictor, clustered tightly around $\pi/2$).
\item True linear predictor $\eta_i = \beta_0+\beta_1\cos(\theta_i)+\beta_2\sin(\theta_i)$, with $(\beta_0,\beta_1,\beta_2)=(1,2,3)$ and the logit link generating the true success probability, $\pi_i = \exp(\eta_i)/[1+\exp(\eta_i)]$.
\item Binary response $Z_i\sim\text{Bernoulli}(\pi_i)$.
\item For each simulated data set, all five link functions were fit by maximum likelihood and the resulting AIC and deviance recorded; replications that produced quasi-complete separation (a degenerate response vector, or non-finite likelihood under any link) were excluded from the corresponding average and tabulated as estimation failures.
\item $B=500$ Monte Carlo replications were run for each concentration level.
\end{itemize}

This design mirrors the structure of the related circular Poisson regression simulation of \citet{tasdan2026}, but adds a second predictor-concentration scenario specifically to probe link-function sensitivity to how widely the circular covariate is spread around the circle, a consideration that has no counterpart in regression with linear predictors.

\subsection*{Results}

\begin{figure}[ht]
\centering
\includegraphics[width=0.62\textwidth]{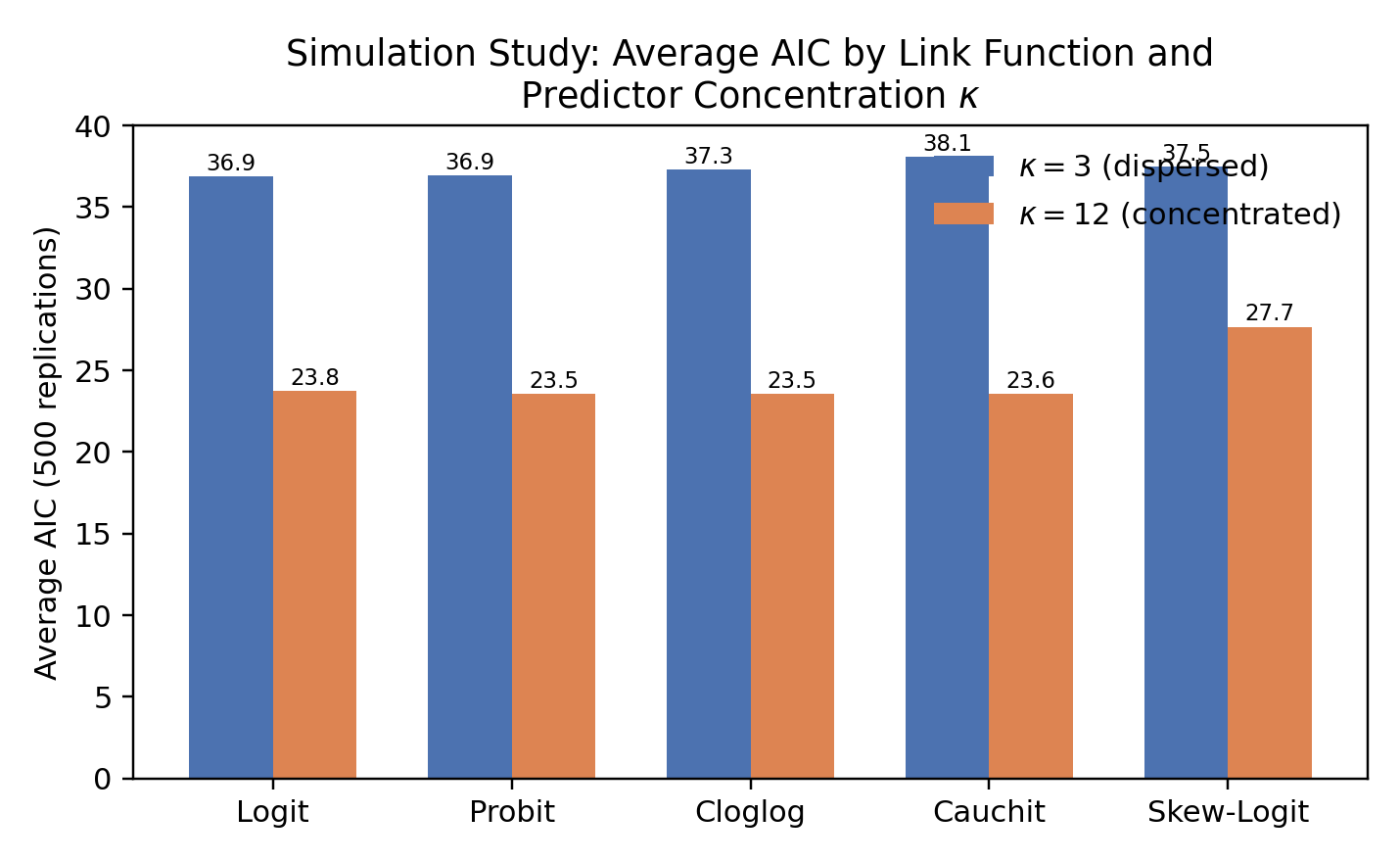}
\caption{Average AIC by link function under dispersed ($\kappa=3$) and concentrated ($\kappa=12$) circular predictors.}
\label{fig:simaic}
\end{figure}

Table~\ref{tab:simresults} summarizes the average AIC across replications for each link function and concentration level (the corresponding deviances follow the same ordering, since all symmetric and the cloglog links use three estimated parameters; the skew-logit link uses four). Figure~\ref{fig:simaic} displays the same comparison graphically.

\begin{table}[ht]
\centering
\caption{Average AIC over 500 Monte Carlo replications, by link function and von Mises concentration $\kappa$ of the circular predictor.}
\label{tab:simresults}
\begin{tabular}{lcc}
\toprule
\textbf{Link function} & $\kappa=3$ (dispersed) Avg.\ AIC & $\kappa=12$ (concentrated) Avg.\ AIC \\
\midrule
Logit & 36.87 & 23.75 \\
Probit & 36.93 & 23.54 \\
Cloglog & 37.31 & 23.55 \\
Cauchit & 38.11 & 23.59 \\
Skew-logit & 37.48 & 27.67 \\
\bottomrule
\end{tabular}
\end{table}

Two patterns emerge. First, under the dispersed predictor ($\kappa=3$), all five links produce broadly comparable average AIC values (36.9--38.1), with logit and probit which are the symmetric links slightly favored on average and the skew-logit link achieving the lowest average deviance despite its extra parameter, reflecting the additional flexibility it offers when the predictor spans a wide arc of the circle. Second, under the concentrated predictor ($\kappa=12$), the four three-parameter links (logit, probit, cloglog, Cauchit) again perform almost identically to one another, but the skew-logit link performs distinctly worse (average AIC 27.67 versus approximately 23.5--23.8 for the others). When $\theta_i$ is tightly clustered around a single direction, $\cos(\theta_i)$ and $\sin(\theta_i)$ vary over only a narrow range and become highly correlated with one another and with the intercept; the extra shape parameter of the skew-logit link is then poorly identified from the data, inflating the AIC penalty without a compensating improvement in fit.

A second, more consequential finding concerns estimability. Under $\kappa=3$, only 1 of 500 replications (0.2\%) failed due to quasi-complete separation. Under $\kappa=12$, however, 53 of 500 replications (10.6\%) failed, because the narrow angular spread of $\theta$ made it possible for the small sample to be perfectly or quasi-perfectly classified by $\cos(\theta_i)$ and $\sin(\theta_i)$ alone. This is a practical caution specific to circular logistic regression that has no direct analogue in ordinary logistic regression with linear covariates: a highly concentrated circular predictor, however precisely measured, carries comparatively little information for discriminating a binary response once it is projected onto $\cos(\theta)$ and $\sin(\theta)$, and the resulting near-collinearity increases the chance of unstable or non-existent maximum likelihood estimates. We return to this point in Section~8.

\section{Real Data Applications}

\subsection{Wind Direction and Rainfall Occurrence in Macomb, Illinois}

Daily values of rainfall occurrence (rain $=1$, no rain $=0$) and the corresponding wind direction were obtained for the city of Macomb, Illinois ($n=103$ days, August 1 to November 11), from the Visual Crossing weather data service. Wind direction was recorded in degrees and converted to radians. Of the 103 days, 44 recorded measurable rainfall and 59 did not, a moderately but not severely unbalanced split ($\approx 43\%$ positive).

All five link functions of Table~\ref{tab:links}, together with skew-Laplace and skew-Cauchy variants fit during the original analysis, were applied to the model $g(\pi_i)=\beta_0+\beta_1\cos(\theta_i)+\beta_2\sin(\theta_i)$, with $\theta_i$ the wind direction in radians. Table~\ref{tab:rainfall} reports the resulting AIC and deviance.

\begin{table}[ht]
\centering
\caption{Model comparison for the Macomb, IL rainfall-occurrence data ($n=103$).}
\label{tab:rainfall}
\begin{tabular}{lcc}
\toprule
\textbf{Link} & \textbf{AIC} & \textbf{Deviance} \\
\midrule
Logit & 61.845 & 55.845 \\
Probit & 61.828 & 55.828 \\
Cloglog & 61.848 & 55.848 \\
Skew-logit & 61.845 & 55.845 \\
Skew-Laplace & 61.852 & 55.852 \\
Skew-Cauchy & 61.912 & 55.912 \\
\bottomrule
\end{tabular}
\end{table}

The probit link achieves the lowest AIC, but the spread across all six links is under 0.1 AIC units, effectively indistinguishable in practice. This is consistent with the simulation finding of Section~6 and with the theoretical guidance of \citet{chen1999}: because the rainfall split (44 vs.\ 59) is only moderately unbalanced, none of the candidate links is rewarded over the others, and the added flexibility of the skew links provides no measurable benefit. Asymmetric links are expected to earn their keep specifically when the response proportions are far more lopsided, or when there is independent subject-matter reason to expect faster approach to one boundary than the other \citep{collett2003,gomezdeniz2022}.

Fitting the skew-logit model in full gives the coefficient estimates summarized in Table~\ref{tab:rainfallcoef} (the dispersion parameter is fixed at 1, as appropriate for the binomial family):

\begin{table}[ht]
\centering
\caption{Fitted circular skew-logit model for the rainfall data.}
\label{tab:rainfallcoef}
\begin{tabular}{lc}
\toprule
\textbf{Term} & \textbf{Estimate} \\
\midrule
Intercept ($\beta_0$) & $-2.531$ \\
$\cos(\theta)$ ($\beta_1$) & $0.041$ \\
$\sin(\theta)$ ($\beta_2$) & $-0.374$ \\
\bottomrule
\end{tabular}
\end{table}

\begin{figure}[ht]
\centering
\includegraphics[width=0.58\textwidth]{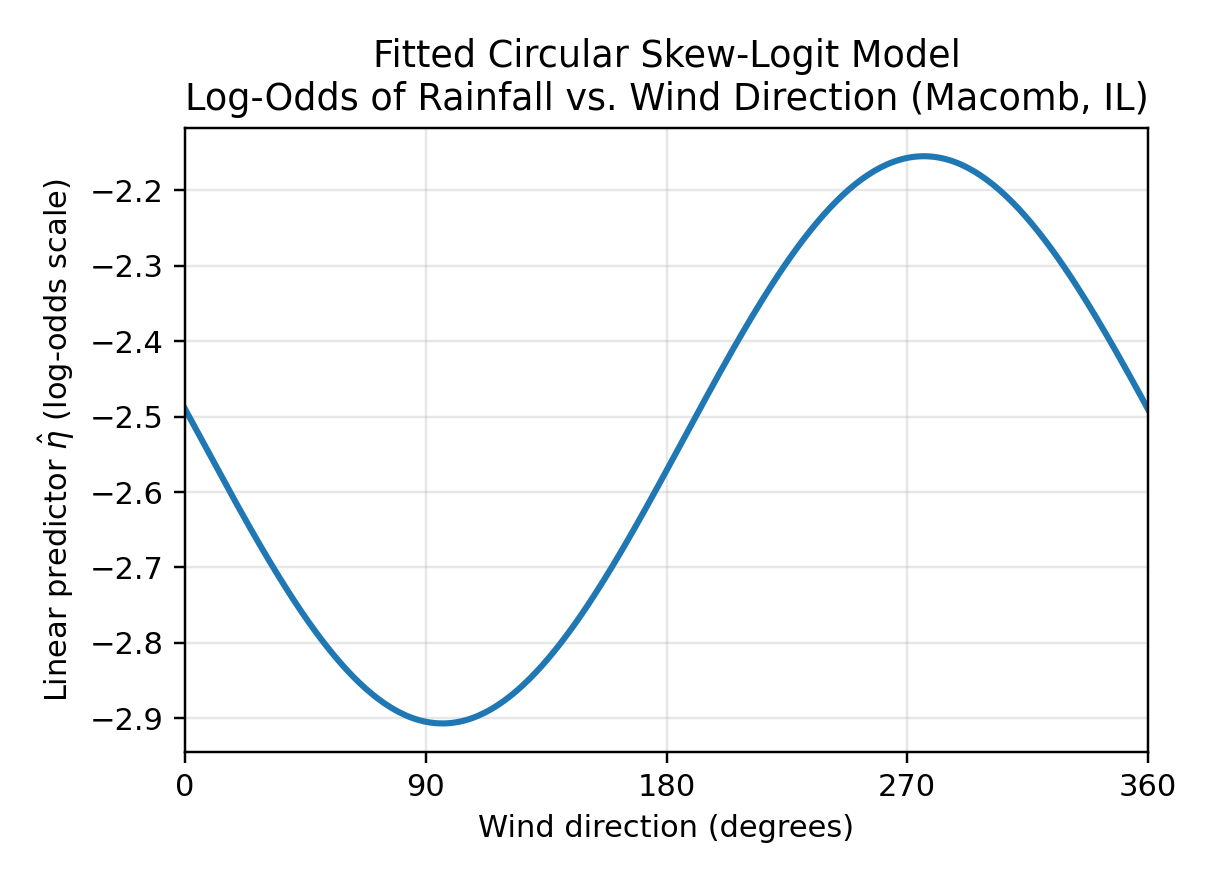}
\caption{Fitted log-odds of rainfall as a function of wind direction, from the circular skew-logit model of Table~\ref{tab:rainfallcoef}.}
\label{fig:rainfallfit}
\end{figure}

From these coefficients, the recovered amplitude and phase angle are $A=\sqrt{\beta_1^2+\beta_2^2}\approx 0.376$ and $\theta_0=\text{atan2}(\beta_2,\beta_1)\approx 276^\circ$. That is, the log-odds of rainfall are estimated to be highest when the wind blows from a direction near $276^\circ$ (west-northwest) and lowest from the roughly opposite direction near $90$--$96^\circ$ (east), a modest but interpretable directional signal (Figure~\ref{fig:rainfallfit}). The negative intercept ($-2.53$ on the logit scale at the model's circular origin) reflects the overall rarity of rainfall on a given day within this observation window.

\subsection{Connecting to Circular Count Data: Earthquake Occurrence by Month}

\begin{figure}[ht]
\centering
\includegraphics[width=0.55\textwidth]{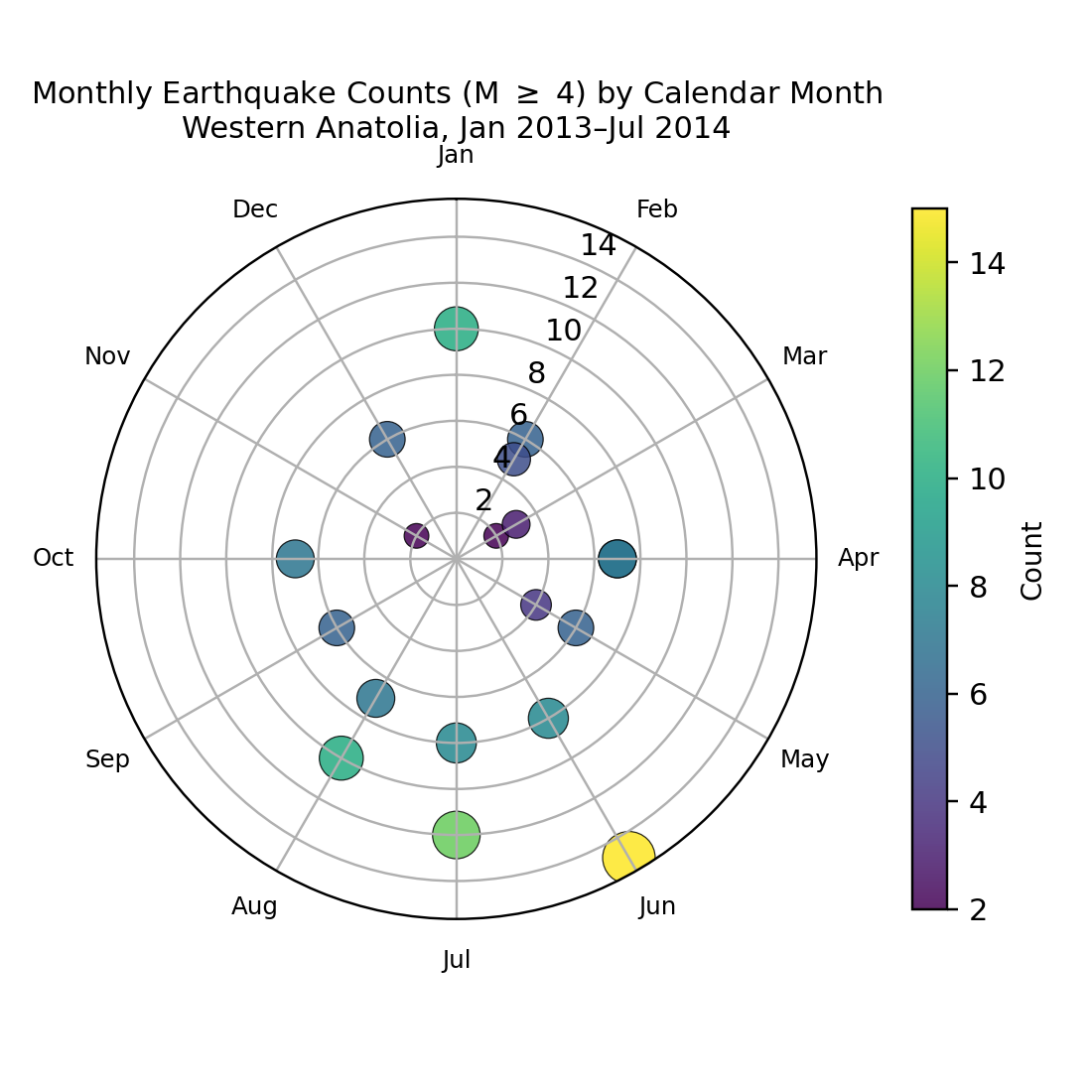}
\caption{Circular plot of monthly earthquake counts by calendar month.}
\label{fig:eqcirc}
\end{figure}

To situate the circular logistic model within the broader family of circular generalized linear models, we revisit the Western Anatolia earthquake data analyzed under a circular Poisson regression model by \citet{tasdan2026}. The data record the monthly count of earthquakes of magnitude 4 or higher in the Aegean region of Turkiye from January 2013 to July 2014, with calendar month converted to an angle via $\theta = \text{month}\times 2\pi/12$ (Table~\ref{tab:eqdata}; Figure~\ref{fig:eqcirc}).

\begin{table}[ht]
\centering
\caption{Monthly earthquake counts ($M\ge4$), Western Anatolia, January 2013--July 2014 \citep{tasdan2026}.}
\label{tab:eqdata}
\small
\begin{tabular}{lcclc}
\toprule
\textbf{Month} & \textbf{Counts} & \quad & \textbf{Month} & \textbf{Counts} \\
\midrule
Jan & 6, 7  & & Jul & 10, 7 \\
Feb & 2, 4  & & Aug & 6 \\
Mar & 7, 8  & & Sep & 7 \\
Apr & 6, 12 & & Oct & 2 \\
May & 15, 7 & & Nov & 6 \\
Jun & 8, 6  & & Dec & 10 \\
\bottomrule
\end{tabular}

\vspace{2pt}
\footnotesize{Note: multiple values list the separate monthly counts recorded for that calendar month across the Jan 2013--Jul 2014 observation window (Table 4 of \citealp{tasdan2026}).}
\end{table}

The circular Poisson regression model fit to these data, $g(\lambda_i)=\log(\lambda_i)=\beta_0+\beta_1\cos(\theta_i)+\beta_2\sin(\theta_i)$, uses the identical cos--sin engineering of the circular predictor as the circular logistic model of Equation~\eqref{eq:circlogit2}, differing only in the response distribution (Poisson rather than Bernoulli/binomial) and the corresponding log link in place of logit, probit, or cloglog. The fitted coefficients were $\beta_0=1.893$ ($p<.001$), $\beta_1=0.365$ ($p=.005$), and $\beta_2=-0.228$ ($p=.062$), giving recovered amplitude $A\approx 0.431$ and phase angle $\theta_0\approx 328^\circ$, indicating that the expected count of earthquakes peaks for months falling near late November and is lowest roughly six months later. The likelihood ratio test against the null (intercept-only) model was highly significant (residual deviance 16.57 on null deviance 27.96, $p<.001$), confirming a genuine circular pattern in seismic occurrence over the calendar year.

This example illustrates that the cosine--sine linear predictor of Equation~\eqref{eq:circlogit2} is not specific to the binomial family: it provides a unifying device for circular regression across exponential-family response types, with the link function (logit/probit/cloglog/Cauchit/skew-logit for binary or binomial data, log for Poisson counts) chosen according to the nature of the response. Developing this unification more formally, a circular GLM framework spanning multiple response families with a common circular-predictor variable, is a natural direction for future work, discussed further in Section~9.

\section{Discussion and Practical Guidelines}

\begin{itemize}
\item \textbf{Match the link to the balance of the response.} When the binary or binomial response is close to balanced (as in the Macomb rainfall data), symmetric links (logit, probit) and asymmetric alternatives perform similarly, and the simpler symmetric link is preferred on parsimony grounds. When the response is markedly unbalanced, the literature \citep{chen1999,collett2003,gomezdeniz2022} and the bioassay comparisons of \citet{dawotola2025} indicate that an asymmetric link such as cloglog or a skew-logit/skew-Cauchy family is likely to fit better.
\item \textbf{Check the concentration of the circular predictor before adding link-function flexibility.} The simulation in Section~6 shows that highly concentrated circular predictors (large $\kappa$) increase the risk of quasi-complete separation and make additional shape parameters (as in the skew-logit link) poorly identified; in such cases, a parsimonious symmetric link is the safer default, and larger sample sizes or a wider observation window for the circular covariate should be sought where feasible.
\item \textbf{Report the recovered amplitude and phase angle.} Beyond the individual $\beta_1,\beta_2$ coefficients, the amplitude $A$ and phase angle $\theta_0$ of Equation~\eqref{eq:circlogit2} carry the directly interpretable circular information (the strength and location, respectively, of the directional effect) and should be reported alongside standard GLM output, as illustrated in both real-data examples.
\item \textbf{Software gap.} None of the principal circular statistics packages in R including \texttt{circular} \citep{agostinelli2017,hornik2014} or the routines collated by \citet{pewsey2013} currently provide a dedicated circular logistic regression routine with a menu of link functions, despite the availability of custom-link machinery within \texttt{glm()} that makes the implementation straightforward once the cosine-sine predictor has been formed. The same gap exists in standard Python GLM software. Closing it, with built-in support for symmetric and skewed links and automatic computation of $A$ and $\theta_0$ with standard errors (e.g., via the delta method), would substantially lower the barrier to applying these methods.
\end{itemize}

\section{Conclusion}

This paper developed a circular logistic regression model for binary and binomial responses observed jointly with a circular predictor, extending the established linear-circular regression framework and paralleling the circular Poisson regression model recently proposed for circular count data \citep{tasdan2026}. We assembled and compared five link functions of logit, probit, complementary log-log, Cauchit, and a skew-logit power links within this framework, motivated by an extensive literature documenting that link misspecification can materially bias regression estimates and degrade model fit \citep{czado1992,chen1999,collett2003,gomezdeniz2022,dawotola2025}.

A Monte Carlo simulation showed that the practical impact of link choice in the circular setting depends critically on the concentration of the circular predictor: with a dispersed predictor, all five links performed comparably, with the added flexibility of the skew-logit link yielding the best average fit; with a concentrated predictor, the three-parameter symmetric links remained stable while the skew-logit link both fit worse on average and failed considerably more often due to quasi-complete separation. Two real data applications, wind direction and rainfall occurrence in Macomb, Illinois, and monthly earthquake counts in Western Anatolia, illustrated the methodology and demonstrated that the same cosine--sine circular-predictor device unifies circular regression across the binomial and Poisson families.

Several extensions are natural next steps: incorporating multiple circular predictors or a mixture of circular and linear covariates; developing a fully general skew-Cauchy and skew-Laplace circular GLM with jointly estimated shape and scale parameters; deriving analytic standard errors for the recovered amplitude $A$ and phase angle $\theta_0$ via the delta method; exploring Bayesian estimation as an alternative to maximum likelihood, particularly for the quasi-separation cases identified in Section~6; and packaging the resulting methods, with automatic link-function selection, as an R or Python software contribution to fill the gap identified in Section~8. We hope this work, together with the companion circular Poisson regression study of \citet{tasdan2026}, helps establish circular logistic regression as a usable and well-understood tool for researchers working with directional predictors and discrete or binary outcomes.


\begin{thebibliography}{}

\bibitem[Agostinelli and Lund, 2017]{agostinelli2017}
Agostinelli, C. and Lund, U. (2017).
\newblock R package ``circular'': Circular statistics.
\newblock CRAN.

\bibitem[Agresti, 2002]{agresti2002}
Agresti, A. (2002).
\newblock {\em Categorical Data Analysis}.
\newblock Wiley, 2nd edition.

\bibitem[Breen et~al., 2018]{breen2018}
Breen, R., Karlson, K.~B., and Holm, A. (2018).
\newblock Interpreting and understanding logits, probits, and other nonlinear
  probability models.
\newblock {\em Annual Review of Sociology}, 44(1):39--54.

\bibitem[Chen et~al., 1999]{chen1999}
Chen, M.-H., Dey, D.~K., and Shao, Q.-M. (1999).
\newblock A new skewed link model for dichotomous quantal response data.
\newblock {\em Journal of the American Statistical Association},
  94(448):1172--1186.

\bibitem[Collett, 2003]{collett2003}
Collett, D. (2003).
\newblock {\em Modelling Binary Data}.
\newblock Chapman \& Hall/CRC, 2nd edition.

\bibitem[Czado and Santner, 1992]{czado1992}
Czado, C. and Santner, T.~J. (1992).
\newblock The effect of link misspecification on binary regression inference.
\newblock {\em Journal of Statistical Planning and Inference}, 33:213--231.

\bibitem[Dawotola and Tasdan, 2025]{dawotola2025}
Dawotola, T.~B. and Tasdan, F. (2025).
\newblock A comparative analysis of some link functions for binomial regression
  models with applications to bioassay data.
\newblock {\em International Journal of Scientific Research and Modern
  Technology}, 3(12):106--116.

\bibitem[Fisher, 1993]{fisher1993}
Fisher, N.~I. (1993).
\newblock {\em Statistical Analysis of Circular Data}.
\newblock Cambridge University Press.

\bibitem[Fisher and Lee, 1992]{fisherlee1992}
Fisher, N.~I. and Lee, A.~J. (1992).
\newblock Regression models for an angular response.
\newblock {\em Biometrics}, 48(3):665--677.

\bibitem[G\'omez-D\'eniz et~al., 2022]{gomezdeniz2022}
G\'omez-D\'eniz, E., Calder\'in-Ojeda, E., and G\'omez, H.~W. (2022).
\newblock Asymmetric versus symmetric binary regression: A new proposal with
  applications.
\newblock {\em Symmetry}, 14(4):733.

\bibitem[Hornik, 2014]{hornik2014}
Hornik, K. (2014).
\newblock R package ``circular'': Circular statistics (version 0.4-93).

\bibitem[Jammalamadaka and SenGupta, 2001]{jammalamadaka2001}
Jammalamadaka, S.~R. and SenGupta, A. (2001).
\newblock {\em Topics in Circular Statistics}.
\newblock World Scientific Publishing.

\bibitem[Kadhem and Khan, 2017]{kadhem2017}
Kadhem, A.-D. and Khan, S. (2017).
\newblock Logistic regression for circular data.
\newblock {\em AIP Conference Proceedings}, 1842(1):030022.

\bibitem[Landler et~al., 2021]{landler2021}
Landler, L., Ruxton, G.~D., and Malkemper, E.~P. (2021).
\newblock Advice on comparing two independent samples of circular data in
  biology.
\newblock {\em Scientific Reports}, 11:20337.

\bibitem[Li, 2014]{li2014}
Li, J. (2014).
\newblock {\em Choosing the Proper Link Function for Binary Data}.
\newblock PhD thesis, The University of Texas at Austin.

\bibitem[Mardia and Jupp, 2009]{mardia2009}
Mardia, K.~V. and Jupp, P.~E. (2009).
\newblock {\em Directional Statistics}.
\newblock Wiley.

\bibitem[McCullagh and Nelder, 1989]{mccullagh1989}
McCullagh, P. and Nelder, J.~A. (1989).
\newblock {\em Generalized Linear Models}.
\newblock Chapman \& Hall/CRC, 2nd edition.

\bibitem[Pewsey et~al., 2013]{pewsey2013}
Pewsey, A., Neuhauser, M., and Ruxton, G.~D. (2013).
\newblock {\em Circular Statistics in R}.
\newblock Oxford University Press.

\bibitem[Rivest et~al., 2015]{rivest2015}
Rivest, L.-P., Duchesne, T., Nicosia, A., and Fortin, D. (2015).
\newblock A general angular regression model for the analysis of data on animal
  movement in ecology.
\newblock {\em Journal of the Royal Statistical Society: Series C},
  64(3):445--463.

\bibitem[Shim et~al., 2023]{shim2023}
Shim, H., Bonifay, W., and Wiedermann, W. (2023).
\newblock Parsimonious asymmetric item response theory modeling with the
  complementary log-log link.
\newblock {\em Behavior Research Methods}, 55(1):200--219.

\bibitem[Smithson and Shou, 2017]{smithson2017}
Smithson, M. and Shou, Y. (2017).
\newblock Cdf-quantile distributions for modelling random variables on the unit
  interval.
\newblock {\em British Journal of Mathematical and Statistical Psychology},
  70(3):412--438.

\bibitem[Tasdan, 2026]{tasdan2026}
Tasdan, F. (2026).
\newblock Modelling of count data in circular statistics.
\newblock In Duman, O. and Erkus-Duman, E., editors, {\em Approximation Theory
  and Special Functions}, volume 503 of {\em Springer Proceedings in
  Mathematics \& Statistics}, pages 593--602. Springer Nature.

\bibitem[Tasdan and Cetin, 2013]{tasdan2013}
Tasdan, F. and Cetin, M. (2013).
\newblock A simulation study on the influence of ties on uniform scores test
  for circular data.
\newblock {\em Journal of Applied Statistics}, 41(5):1137--1146.

\bibitem[Tasdan and Yeniay, 2014]{tasdan2014}
Tasdan, F. and Yeniay, O. (2014).
\newblock Power study of circular anova test against nonparametric
  alternatives.
\newblock {\em Hacettepe Journal of Mathematics and Statistics}, 43(1):97--115.

\bibitem[Tasdan and Yeniay, 2018]{tasdan2018}
Tasdan, F. and Yeniay, O. (2018).
\newblock A comparative simulation of multiple testing procedures in circular
  data problems.
\newblock {\em Journal of Applied Statistics}, 45(2):255--269.

\bibitem[Zaidi and Al~Luhayb, 2023]{zaidi2023}
Zaidi, A. and Al~Luhayb, A. S.~M. (2023).
\newblock Two statistical approaches to justify the use of the logistic
  function in binary logistic regression.
\newblock {\em Mathematical Problems in Engineering}, 2023:5525675.

\end{thebibliography}
\end{document}